\documentclass
[superscriptaddress,secnumarabic,amssymb,amsmath,nobibnotes,aps,prd,showkeys,showpacs,nofootinbib,onecolumn,12pt]{revtex4}%
\usepackage{graphicx}
\usepackage{epsf}
\usepackage{bm}
\usepackage{amsmath}
\usepackage{amsfonts}
\usepackage{amssymb}%
\providecommand{\U}[1]{\protect\rule{.1in}{.1in}}

\newcommand{\be}{\begin{equation}}
\newcommand{\ee}{\end{equation}}

\newcommand{\mincir}{\raise
-3.truept\hbox{\rlap{\hbox{$\sim$}}\raise4.truept\hbox{$<$}\ }}
\newcommand{\magcir}{\raise
-3.truept\hbox{\rlap{\hbox{$\sim$}}\raise4.truept\hbox{$>$}\ }}

\begin{document}
\title{$f(T)$ Cosmology with Nonzero Curvature}
\author{Andronikos Paliathanasis}
\email{anpaliat@phys.uoa.gr}
\affiliation{Institute of Systems Science, Durban University of Technology, Durban 4000,
South Africa }
\affiliation{Instituto de Ciencias F\'{\i}sicas y Matem\'{a}ticas, Universidad Austral de
Chile, Valdivia 5090000, Chile}

\begin{abstract}
We investigate exact and analytic solutions in $f\left(  T\right)  $ gravity
within the context of a Friedmann--Lema\^{\i}tre--Robertson--Walker background
space with nonzero spatial curvature. For the power law theory $f\left(
T\right)  =T^{n}$ we find that the field equations admit an exact solution
with a linear scalar factor for negative and positive spatial curvature. That
Milne-like solution is asymptotic behaviour for the scale factor near the
initial singularity for the model $f\left(  T\right)  =T+f_{0}T^{n}-2\Lambda$.
The analytic solution for that specific theory is presented in terms of
Painlev\'{e} Series for $n>1$. Moreover, from the\ value of the resonances of
the Painlev\'{e} Series we conclude that the Milne-like solution is always
unstable while for large values of the indepedent parameter, the field
equations provide an expanding universe with a de Sitter expansion of a
positive cosmological constant. Finally, the presence of the cosmological term
$\Lambda$ in the studied $f\left(  T\right)  $ model plays no role in the
general behavior of the cosmological solution and the universe immerge in a de
Sitter expansion either when the cosmological constant term $\Lambda$ in the
$f\left(  T\right)  $ model vanishes.

\end{abstract}
\date{\today}
\maketitle

\section{Introduction}

\label{sec1}

Teleparallel theory of gravity \cite{ein28,cc} and its modifications
\cite{fer1,fer2,fer3,fer4,fer5,fer6,fer7,myr11} have attracted the attention
of cosmologists over the last year because it can provide a geometric
explanation for the explanation of the recent observations \cite{sp1,fg5}, for
a recent review we refer the reader to \cite{rew1}. The modified teleparallel
theories of gravity belong to the family of theories in the Lorentz symmetry
is violated \cite{li1,lv1}. In teleparallelism, the fundamental connection is
the curvature-less Weitzenb{\"{o}}ck connection \cite{Weitzenb23} while the
torsion scalar $T$ is used for the definition of the gravitational Action
Integral \cite{Hayashi79}. In contrary to General Relativity in which the
Levi-Civita connection and the Ricciscalar $R$ are the fundamental geometric
objects of the theory.

When the gravitational Action Integral is linear on the torsion scalar $T$,
then the theory is equivalent to General Relativity and it is know as TEGR.
However, there are various originally proposed gravitational Lagrangians such
is the Teleparallel dark energy theory, the $f\left(  T\right)  $ theory and
its extensions, see for instance
\cite{fer5,fer6,fer7,myr11,ss1,ss2,mm1,mm2,mm3,mm4}.

In this study we consider the $f\left(  T\right)  $ gravity in a
Friedmann--Lema\^{\i}tre--Robertson--Walker background space. The theory has
been proposed originally as a geometric dark energy candidate in a spatially
flat background space \cite{fer1}. However, during the last years it has been
found that it can explain various eras of the cosmological history. In
$f\left(  T\right)  $ cosmology, the field equations are of second-order and
they have only one dependent variable, the scale factor of the underlying
geometry. Recently, in \cite{ftcur} the case of nonzero spatial curvature for
the background space has been considered. The authors investigated the
existence of bounces and static solutions, while the conditions for the
existence of these solutions were investigated.

We focus on the existence of exact and analytic solutions for the cosmological
field equations in $f\left(  T\right)  $ cosmology where the spatial curvature
is nonzero. A background space with nonzero spatial curvature is not excluded
by the inflationary scenario \cite{in1,in1a}. Indeed, inflation is immune to
negative curvature while the energy density which corresponds to the curvature
can be nonzero in the pre-inflationary era \cite{in2,in3}. Moreover, we shall
focus on the existence of the de Sitter expansion as described by a positive
cosmological constant. Such an expansion is necessary because it provides a
rapid expansion for the size of the universe such that the latter effectively
loses its memory on the initial conditions, which means that the de Sitter
expansion solves the \textquotedblleft flatness\textquotedblright,
\textquotedblleft horizon\textquotedblright\ and monopole problem
\cite{f1,f2}. The cosmic \textquotedblleft no-hair\textquotedblright%
\ conjecture states that all expanding universes with a positive cosmological
constant admit as an asymptotic solution the de Sitter universe \cite{nh1,nh2}%
. The plan of the paper is as follows.

In Section \ref{sec2}, we present the field equations for the cosmological
model of our consideration. In Section \ref{sec3} we consider the power-law
$f\left(  T\right)  =T^{n}$ theory where we find a generalized Milne
solution\ that exists for the field equations for negative and positive
spatial curvature. A more general function $f\left(  T\right)  $ is considered
in Section \ref{sec4}. Specifically we select the $f\left(  T\right)
=T+f_{0}T^{n}-2\Lambda$ model, which has been proposed as a dark energy
candidate. We write the field equations where we find that they admit a
movable singularity. Near the singularity and for $n>1$ the field equations
are dominated by the $T^{n}$ term which means that the generalized Milne
solution describes the scale factor at the singularity. Furthermore, we study
if the field equations admit the Painlev\'{e} property. We find that for $n>1$
the field equations pass the Painlev\'{e} test and the resonances give that
the analytic solution can be expressed by a Right Painlev\'{e} Series. The
latter indicates an expanding universe which leads to a de Sitter expansion.
Finally, in Section \ref{sec5} we summarize our results and we draw our conclusions.

\section{$f\left(  T\right)  $ Cosmology with nonzero curvature}

\label{sec2}

The modified teleparallel modified theory of gravity known as $f\left(
T\right)  $ theory is a second-order theory which violates the Lorentz
symmetry for any nonlinear function $f$~~\cite{li1}. The fundamental geometric
objects of the theory are the vierbein fields $e^{i}(x^{k})$ which define the
unholonomic frame of the theory. For the vierbein fields it follows
$g(e_{i},e_{j})=e_{i}.e_{j}=\eta_{ij}$, while in a coordinate system
$e^{i}(x^{k})=h_{\mu}^{i}(x^{k})dx^{i}$ such that $g_{\mu\nu}(x^{k})=\eta
_{ij}h_{\mu}^{i}(x^{k})h_{\nu}^{j}(x^{k})$ where $h_{\mu}^{i}$ is the dual
basis of the theory. The invariant which is used for the definition of the
Lagrangian in teleparallelism is the scalar $T$ for the curvatureless
Weitzenb\"{o}ck connection \cite{Weitzenb23}. Because of the existence of the
unholonomic tensor the non-null torsion is defined as
\begin{equation}
T_{\mu\nu}^{\beta}=\hat{\Gamma}_{\nu\mu}^{\beta}-\hat{\Gamma}_{\mu\nu}^{\beta
}=h_{i}^{\beta}(\partial_{\mu}h_{\nu}^{i}-\partial_{\nu}h_{\mu}^{i})\;.
\label{Wein}%
\end{equation}
where the scalar $T$ is given by the expression%
\begin{equation}
T=\frac{1}{2}(K_{\beta}^{\mu\nu}+\delta_{\beta}^{\mu}T_{\theta}^{\theta\nu
}-\delta_{\beta}^{\nu}T_{\theta}^{\theta\mu})T_{\mu\nu}^{\beta}
\label{lagrangian}%
\end{equation}
where $K_{\beta}^{\mu\nu}$\ equals the difference of the Levi Civita
connection in the holonomic and the unholonomic frame and it is defined$~$%
as$\ K_{\beta}^{\mu\nu}=-\frac{1}{2}(T_{\beta}^{\mu\nu}-T_{\beta}^{\nu\mu
}-T_{\beta}^{\mu\nu}).$

The Action Integral in $f\left(  T\right)  $ theory is defined as
\begin{equation}
S_{f\left(  T\right)  }=\frac{1}{16\pi G}\int{d^{4}xef(T)+S}_{m},
\label{action11}%
\end{equation}
in which ${S}_{m}$ is the Action Integral of the matter source and $G$ is
Newton's constant. In the case where $f\left(  T\right)  $ is a linear
function, the teleparallel equivalence of General relativity, with or without
the cosmological constant, is recovered.

Variation with respect to the vierbein fields of the gravitational Action
Integral (\ref{action11}) provides the field equations \cite{fer1}
\begin{align}
&  e^{-1}\partial_{\mu}(e{S}_{i}^{\mu\nu})f^{\prime}(T)-h_{i}^{\lambda}%
T_{\mu\lambda}^{\beta}S_{\beta}^{\nu\mu}f^{\prime}(T)\nonumber\\
&  +S_{i}^{\mu\nu}\partial_{\mu}(T)f^{\prime\prime}(T)+\frac{1}{4}h_{i}^{\nu
}f(T)=4\pi Gh_{i}^{\beta}T_{\beta}^{\nu} \label{equations}%
\end{align}
in which $f^{\prime}\left(  T\right)  =\frac{df\left(  T\right)  }{dT}$,
$f^{\prime\prime}\left(  T\right)  =\frac{d^{2}f}{dT^{2}}$, $T_{\mu\nu}$
includes the contribution of the matter source in the field equation and
${S_{i}}^{\mu\nu}={h_{i}}^{\beta}S_{\beta}^{\mu\nu}$ where $S_{\beta}^{\mu\nu
}=\frac{1}{2}(K_{\beta}^{\mu\nu}+\delta_{\beta}^{\mu}T_{\theta}^{\theta\nu
}-\delta_{\beta}^{\nu}T_{\theta}^{\theta\mu})$.

In the case of a spatially flat Friedmann--Lema\^{\i}tre--Robertson--Walker
(FLRW) universe
\begin{equation}
ds^{2}=dt^{2}-a^{2}\left(  t\right)  \left(  dx^{2}+dy^{2}+dz^{2}\right)  ,
\label{fc.01}%
\end{equation}
then when we select the diagonal unholonomic frame $e^{A}=\left(  dt,a\left(
t\right)  dx,a\left(  t\right)  dy,a\left(  t\right)  dz\right)  $, the field
equations (\ref{equations}) are in agreement with the Action integral
(\ref{action11}).

However, when we consider a nonzero spatial curvature,$~K\neq0,$ in the FLRW
universe,%
\begin{equation}
ds^{2}=dt^{2}-a^{2}\left(  t\right)  \left(  dr^{2}+\sin^{2}\left(
\phi\right)  \left(  d\theta^{2}+\sin^{2}\theta d\phi^{2}\right)  \right)
,~K=1~, \label{fc.02}%
\end{equation}%
\begin{equation}
ds^{2}=dt^{2}-a^{2}\left(  t\right)  \left(  dr^{2}+\sinh^{2}\left(
\phi\right)  \left(  d\theta^{2}+\sin^{2}\theta d\phi^{2}\right)  \right)
,~K=-1~,
\end{equation}
a nondiagonal frame should be considered. Indeed, the selected unholonomic
frame is $e_{i}^{A}=\left(  dt,a\left(  t\right)  E^{r}\left(  K\right)
,a\left(  t\right)  E^{\theta}\left(  K\right)  ,a\left(  t\right)  E^{\phi
}\left(  K\right)  \right)  $, in which for $K=1$, \cite{ff1}
\begin{equation}
E^{r}\left(  K=1\right)  =-\cos\theta dr+\sin r\sin\theta\left(  \cos
rd\theta-\sin r\sin\theta d\phi\right)  ~, \label{fc.03}%
\end{equation}%
\begin{align}
E^{\theta}\left(  K=1\right)   &  =\sin\theta\cos\phi dr-\sin r\left(  \sin
r\sin\phi-\cos r\cos\theta\cos\phi\right)  d\theta+\nonumber\\
&  -\sin r\sin\theta\left(  \cos r\sin\phi+\sin r\cos\theta\cos\phi\right)
d\phi~, \label{fc.04}%
\end{align}%
\begin{align}
E^{\phi}\left(  K=1\right)   &  =-\sin\theta\sin\phi dr-\sin r\left(  \sin
r\cos\phi+\cos r\cos\theta\sin\phi\right)  d\theta+\nonumber\\
&  -\sin r\sin\theta\left(  \cos r\cos\phi-\sin r\cos\theta\sin\phi\right)
d\phi. \label{fc.05}%
\end{align}
On the other hand, for $K=-1,~E^{r}\left(  K\right)  ,~E^{\theta}\left(
K\right)  $ and $E^{\phi}\left(  K\right)  $ are defined as%
\begin{equation}
E^{r}\left(  K=-1\right)  =\cos\theta dr+\sinh r\sin\theta\left(  -\cosh
rd\theta+i\sinh r\sin\theta d\phi\right)  ~, \label{fc.06}%
\end{equation}%
\begin{align}
E^{\theta}\left(  K=-1\right)   &  =-\sin\theta\cos\phi dr+\sinh r\left(
i\sinh r\sin\phi-\cos r\cos\theta\cos\phi\right)  d\theta+\nonumber\\
&  +\sinh r\sin\phi\left(  \cosh r\sin\phi+i\sinh r\cos\theta\cos\phi\right)
d\phi~, \label{fc.07}%
\end{align}%
\begin{align}
E^{\phi}\left(  K=-1\right)   &  =\sin\theta\sin\phi dr+\sinh r\left(  i\sinh
r\cos\phi+\cosh r\cos\theta\sin\phi\right)  d\theta+\nonumber\\
&  +\sinh r\sin\theta\left(  \cosh r\cos\phi-\sinh r\cos\theta\sin\phi\right)
d\phi~. \label{fc.08}%
\end{align}

For this frame, the scalar $T$~is derived%
\begin{equation}
T=6\left(  \frac{K}{a^{2}}-H^{2}\right)  ,~H=\frac{\dot{a}}{a}~,~\dot{a}%
=\frac{da}{dt} \label{fc.10}%
\end{equation}
while the field equations (\ref{equations}) for a isotropic fluid source with
energy density $\rho$ and pressure $p$ are \cite{ff1}%
\begin{equation}
6\left(  H^{2}+\frac{K}{a^{2}}\right)  f^{\prime}+\left(  f\left(  T\right)
-Tf^{\prime}\right)  =2\rho~, \label{fc.11}%
\end{equation}%
\begin{equation}
-4f^{\prime}\left(  2\dot{H}+3H^{2}\right)  +4\left(  \dot{H}+Ka^{-2}\right)
\left(  12H^{2}f^{\prime\prime}+f^{\prime}\right)  -f\left(  T\right)  =2p~,
\label{fc.12}%
\end{equation}
where we observe that of linear function $f\left(  T\right)  $, the usual
Friedmann's equations are recovered.

We continue with the investigation of analytic and exact solutions for the
field equations (\ref{fc.11}), (\ref{fc.12}) for functional forms of $f\left(
T\right)  $ of specific interests.

\section{Generalized Milne universe}

\label{sec3}

In the following we assume that the spacetime is vacuum, that is,~$\rho=0$ and
$p=0$. Therefore, gravitational field equations become%
\begin{equation}
6\left(  H^{2}+\frac{K}{a^{2}}\right)  f^{\prime}+\left(  f\left(  T\right)
-Tf^{\prime}\right)  =0~, \label{fc.14}%
\end{equation}%
\begin{equation}
-4f^{\prime}\left(  2\dot{H}+3H^{2}\right)  +4\left(  \dot{H}+Ka^{-2}\right)
\left(  12H^{2}f^{\prime\prime}+f^{\prime}\right)  -f\left(  T\right)  =0.
\label{fc.15}%
\end{equation}

From the latter, we are able to define the effective energy density for the
geometric fluid to be $\rho_{f\left(  T\right)  }=-\frac{\left(  f\left(
T\right)  -Tf^{\prime}\right)  }{2f^{\prime}}$.

In the following we consider the power-law function $f\left(  T\right)
=T^{n}~$\cite{ls1},~and we investigate the existence of Milne-like solutions.

Indeed for the power-law $f\left(  T\right)  $ function, the field equation
(\ref{fc.14}), (\ref{fc.15}) are written as follows
\begin{equation}
a^{-2n}\left(  K-\dot{a}^{2}\right)  ^{n-1}\left(  K+\left(  2n-1\right)
\dot{a}^{2}\right)  =0~,
\end{equation}%
\begin{equation}
a^{-2n}\left(  K-\dot{a}^{2}\right)  ^{n-2}\left(  \left(  2n-3\right)
\left(  K-\dot{a}^{2}\right)  \left(  K+\left(  2n-1\right)  \dot{a}%
^{2}\right)  -2na\left(  K-\left(  2n-1\right)  \dot{a}^{2}\right)  \ddot
{a}\right)  =0~.
\end{equation}

Hence, for $n>1$, we have the (positive scale factor) solutions $a_{A}\left(
t\right)  =\sqrt{K}t$ and $a_{B}\left(  t\right)  =\sqrt{\frac{K}{1-2n}}t$.
The first solution, namely $a_{A}\left(  t\right)  ,$ is a real function when
$K>0,$ while $a_{B}\left(  t\right)  $ is a real solution and physically
accepted when $K<0$. \ The second solution $a_{B}\left(  t\right)  $ describes
the generalized Milne universe, since $K<0$, where the geometric terms which
follow by $f\left(  T\right)  $ in the field equations mimic the curvature
filed such that to modify the effective curvature from $K\rightarrow\frac
{K}{1-2n}$.\ Moreover, $a_{A}\left(  t\right)  $ is a new solution for
positive curvature where the geometric fluid mimics the curvature term with a
negative effective energy density.

Conversely, $n<1$, there exists only the generalized Milne solution
$a_{B}\left(  t\right)  ~$but in this case the solution exists for $K>0$.
These solutions are the general solutions for the field equations.

It is important to mention that in contrary to the spatially flat FLRW
universe in which the vacuum solution in $f\left(  T\right)  $ gravity is the
Minkowski universe \cite{ftb1}, that is, the vacuum solution of General
Relativity; in the presence of the curvature term the vacuum solutions are
different from that of General\ Relativity.

\section{Analytic solutions}

\label{sec4}

We proceed by considering the more general functions $f\left(  T\right)  $
with special interests in cosmology and astrophysics. Specifically, for the
function $f\left(  T\right)  $ we consider~the dark energy model $f\left(
T\right)  =T+f_{0}T^{n}$~known as dark torsion \cite{ls2}, and the case of the
dark torsion model with the cosmological constant $f\left(  T\right)
=T+f_{0}T^{n}-2\Lambda~$which has been constraint by local gravitational
systems \cite{ls3}.

For these two models, we observe that Milne or Milne-like exact solutions do
not exist as for the power-law theory. Thus, other mathematical techniques
should be applied for the derivation of solutions. We follow the approach
applied in \cite{ftb1} on the application of the singularity analysis for the
construction of analytic cosmological solutions.

The necessary property in order for the singularity analysis to work is the
existence of a movable pole for the differential equation. Hence, we consider
$a\left(  t\right)  =\phi_{0}\left(  t\right)  \phi\left(  t\right)  ^{p}$,
where $\phi\left(  t\right)  _{|t\rightarrow t_{0}}=0$, and $p$ should be a
negative rational number. Parameter $p$ and function $\phi_{0}\left(
t\right)  $ are derived by choosing the master equation of our study, i.e.
equation (\ref{fc.15}) to be dominated by the singular terms. This is known as
the first step in the Ablowitz-Ramani-Segur (ARS) algorithm
\cite{Abl1,Abl2,Abl3}. In singularity analysis the solution is expressed in
terms of Painlev\'{e} Series, thus it is necessary to find the integration
constants and the step of the Painlev\'{e} Series as we move far from the
singularity .

The second step of the algorithm is based on the determination of the
resonances which provide information about the position of the integration
constants of the solution \cite{buntis}. Hence, for the leading order
behaviour\ $a_{l}\left(  t\right)  ~$that we have found, we replace $a\left(
t\right)  =a_{l}\left(  t\right)  \left(  1+\varepsilon\phi\left(  t\right)
^{S}\right)  $ and we linearize around $\varepsilon^{2}\rightarrow0$. From the
leading-order terms of the latter equation we extract a polynomial where by
assuming that it is zero, we find the values of the resonances $S$. The number
of independent values of $S$ should be equal to the order of the differential
equation, since as we mentioned before the resonances are related with the
integration of constants of the solution. Furthermore, the resonance $S=-1$
should exist in order for the singularity to be a simple pole. The third and
final step of the ARS algorithm, known as consistency test, describes how to
write the Painlev\'{e} Series and replace it in the differential equation in
order to test that it is a solution. For more details on the ARS algorithm and
for an extended discussion we refer the reader to \cite{leachan}.

\subsection{Dark torsion$~f\left(  T\right)  =T+f_{0}T^{n}$}

For the dark torsion model with $f\left(  T\right)  =T+f_{0}T^{n}$, the
modified Friedmann equations become%
\begin{equation}
0=12H^{2}\left(  1+nf_{0}T^{n-1}\right)  +T+f_{0}T^{n}~, \label{fc.16}%
\end{equation}%
\begin{align}
0  &  =-4\left(  1+nf_{0}T^{n-1}\right)  \left(  2\dot{H}+3H^{2}\right)
-T-f_{0}T^{n}\nonumber\\
&  +4\left(  1+nf_{0}T^{n-1}\left(  12\left(  n-1\right)  H^{2}+T\right)
\right)  \left(  \dot{H}+Ka^{-2}\right)  ~. \label{fc.17}%
\end{align}

We replace scalar $T$ from (\ref{fc.10}), and we assume that the leading-order
behaviour is described by the $a\left(  t\right)  =\phi_{0}\left(  t\right)
\phi\left(  t\right)  ^{p}$. Usually, $\phi\left(  t\right)  $ is assumed to
be the linear function $\phi\left(  t\right)  =\left(  t-t_{0}\right)  $,
however that it is not necessary \cite{b5}.

For $n>1,$ from equation (\ref{fc.17}) we find that the leading-order terms is
for $p=1$. However,~$p=1,$ it is not acceptable because such value for $p$
does not provide a singular behaviour. In order to overpass that we follow the
approach described in \cite{leachan} and we consider the new variable
$b\left(  t\right)  =\left(  a\left(  t\right)  \right)  ^{-1}$.

Hence, in the new variable we find that the leading-order behaviour of
equation (\ref{fc.17}) is $b\left(  t\right)  =\Phi_{0}\left(  t\right)
\Phi\left(  t\right)  ^{p}$, where $p=-1$ and $K=\left(  \frac{\dot{\Phi
}\left(  t\right)  }{\Phi_{0}\left(  t\right)  }\right)  ^{2}$ or $K=-\left(
1-2n\right)  \left(  \frac{\dot{\Phi}\left(  t\right)  }{\Phi_{0}\left(
t\right)  }\right)  ^{2}$. In the case where $\Phi\left(  t\right)  $ is a
linear function, then the leading-order behaviour is that of the generalized
Milne solution whish is described by the power-law term $T^{n}$ of the model.
From the second step of the ARS algorithm we find the resonances $S=-1$ and
$S=1$, that is, the analytic solution is expressed in terms of the
Painlev\'{e} Series
\begin{equation}
b\left(  t\right)  =\Phi_{0}\left(  t\right)  \left(  \Phi\left(  t\right)
\right)  ^{-1}+\Phi_{1}\left(  t\right)  +\Phi_{2}\left(  t\right)
\Phi\left(  t\right)  +\Phi_{3}\left(  t\right)  \left(  \Phi\left(  t\right)
\right)  ^{2}+...~. \label{fc.18}%
\end{equation}
In order to perform the consistency test we select a specific value for the
exponent $n$. Consider that $n=2$, then by replacing (\ref{fc.18}) in
(\ref{fc.17}) we end up with the following constraint equations.

For positive value $K$, i.e. for the leading-order behaviour with $K=\left(
\frac{\dot{\Phi}\left(  t\right)  }{\Phi_{0}\left(  t\right)  }\right)  ^{2}$
we find that $\Phi_{1}\left(  t\right)  $ is an arbitrary function, and%
\begin{equation}
\Phi_{2}\left(  t\right)  =\frac{\left(  216f_{0}K\left(  \Phi_{1}\right)
^{2}-72f_{0}\sqrt{K}\dot{\Phi}_{1}-1\right)  \dot{\Phi}^{2}+72f_{0}\ddot{\Phi
}+24f_{0}\dot{\Phi}\left(  9\sqrt{K}\Phi_{1}\ddot{\Phi}-\Phi^{\left(
3\right)  }\right)  }{72f_{0}\sqrt{K}\dot{\Phi}^{3}}~,~\text{etc.~.}
\label{fc.19}%
\end{equation}
Moreover, from the constraint equation (\ref{fc.16}) it follows that
\begin{equation}
\Phi_{1}\left(  t\right)  =-\frac{1}{2\sqrt{K}}\frac{\ddot{\Phi}}{\dot{\Phi}}.
\label{fc.20}%
\end{equation}

For negative curvature and for $K=-3\left(  \frac{\dot{\Phi}\left(  t\right)
}{\Phi_{0}\left(  t\right)  }\right)  ^{2}$ we apply the same procedure and we
find the constraint equation for the integration function
\begin{equation}
\Phi_{1}\left(  t\right)  =-\frac{1}{2}\sqrt{-\frac{3}{K}}\frac{\ddot{\Phi}%
}{\dot{\Phi}} \label{fc.21}%
\end{equation}
and%
\begin{equation}
\Phi_{2}\left(  t\right)  =\frac{\left(  1-24f_{0}\left(  5K\Phi_{1}%
^{2}+3\sqrt{-3K}\dot{\Phi}_{1}\right)  \right)  \dot{\Phi}^{2}+144f_{0}%
\ddot{\Phi}+24f_{0}\dot{\Phi}\left(  5\sqrt{-3K}\Phi_{1}\ddot{\Phi}%
-3\Phi^{\left(  3\right)  }\right)  }{72f_{0}\sqrt{-3K}\dot{\Phi}^{3}%
}~,~\text{etc.~.} \label{fc.22}%
\end{equation}

On the other hand, for $n<1$ $\ $we replace $b\left(  t\right)  =\left(
a\left(  t\right)  \right)  ^{-1}$ and we find the leading-order behaviour
$b\left(  t\right)  =\sqrt{\frac{1-2n}{K}}\Phi\left(  t\right)  ^{-1}$, which
is valid for $K>0$, when $\Phi\left(  t\right)  $ is a real function. We apply
the same procedure as before and we find the resonances to be $S=-1$ and
$S=2n-3$, that is, the second resonance is negative which indicates that the
analytic solution is expressed in terms of the Left Painlev\'{e} Series%
\begin{equation}
b\left(  t\right)  =\Phi_{0}\left(  t\right)  \left(  \Phi\left(  t\right)
\right)  ^{-1}+\Phi_{1}\left(  t\right)  \left(  \Phi\left(  t\right)
\right)  ^{-2}+\Phi_{2}\left(  t\right)  \left(  \Phi\left(  t\right)
\right)  ^{-3}+\Phi_{3}\left(  t\right)  \left(  \Phi\left(  t\right)
\right)  ^{-4}+...~. \label{fc.23}%
\end{equation}

However, by replacing in the field equations the latter solution we find that
the latter Painlev\'{e} Series does not solve the differential equations,
thus, the singularity analysis fails.

Let us assume now the case with linear function $\Phi\left(  t\right)
=\left(  t-t_{0}\right)  $ where the leading-order behaviour is that of the
generalized Milne universe, and let us write the analytic solutions. For
$n>1$, the analytic solution is expressed in terms of the Puiseux Series%
\begin{equation}
b\left(  t\right)  =b_{0}\left(  t-t_{0}\right)  ^{-1}+b_{1}+b_{2}\left(
t-t_{0}\right)  +b_{3}\left(  t-t_{0}\right)  ^{2}+...~. \label{fc.24}%
\end{equation}
Hence, for $n=2\,,~$and $K=1$ we find the coefficients%
\begin{equation}
b_{0}=1~,~b_{1}=0~,~b_{2}=-\frac{1}{72f_{0}}~,~b_{3}=0~,~b_{4}=\frac
{1}{3240f_{0}^{2}}~,~b_{5}=0~,~\text{etc.~.}%
\end{equation}
On the other hand for $K=-1~$it follows%
\begin{equation}
b_{0}=\sqrt{3}~,~b_{1}=0~,~b_{2}=\frac{1}{27\sqrt{3}f_{0}}~,~b_{3}%
=0~,~b_{4}=\frac{17}{77760\sqrt{3}f_{0}^{2}}~,~b_{5}=0~,~\text{etc.~.}%
\end{equation}
Someone can use another function $\Phi\left(  t\right)  $ to write the
analytic solution. However, for every smooth function $\Phi\left(  t\right)  $
near the singularity $\Phi\left(  t\right)  \simeq t-t_{0}$ such that
$\Phi\left(  t_{0}\right)  \simeq0$.

For $n>1$, from (\ref{fc.24}) we observe that as far as we move from the
singularity, then the right parts of the Series dominate, which indicate that
when $t_{0}>0$, the behaviour $\left(  t-t_{0}\right)  ^{-1}$ is unstable. As
far as the scale factor $a\left(  t\right)  $ is concerned, it can be easily
constructed by using function $b\left(  t\right)  $.

Indeed for $n=2$ and $K=1$ it follows%
\begin{equation}
a\left(  t\right)  =\left(  t-t_{0}\right)  \left(  1+\frac{1}{72f_{0}}\left(
t-t_{0}\right)  ^{2}-\frac{1}{8640f_{0}^{2}}\left(  t-t_{0}\right)
^{4}+...\right)  , \label{fc.25}%
\end{equation}
while for $K=-1$ we calculate the scale factor%
\begin{equation}
a\left(  t\right)  =\frac{\left(  t-t_{0}\right)  }{\sqrt{3}}\left(
1-\frac{1}{216f_{0}}\left(  t-t_{0}\right)  ^{2}-\frac{1}{19440f_{0}^{2}%
}\left(  t-t_{0}\right)  ^{4}+...\right)  . \label{fc.26}%
\end{equation}

\subsection{Dark torsion with cosmological constant $f\left(  T\right)
=T+f_{0}T^{n}-2\Lambda$}

In the presence of the cosmological constant in the dark torsion model, the
singularity analysis provides the same results as before, with $\Lambda=0$.
The only difference is in the values of the coefficients of the analytic
solution where the cosmological constant is introduced. In order to make it
clear we present the solution for the scale factor $a\left(  t\right)  $ for
$n=2$ and $K=1,~K=-1$ as before.

For $K=1$, the scale factor is
\begin{equation}
a\left(  t\right)  =\left(  t-t_{0}\right)  \left(  1+\frac{1}{72f_{0}}\left(
t-t_{0}\right)  ^{2}-\frac{1+12f_{0}\Lambda}{8640f_{0}^{2}}\left(
t-t_{0}\right)  ^{4}+...\right)  ,
\end{equation}%
\begin{equation}
a\left(  t\right)  =\frac{\left(  t-t_{0}\right)  }{\sqrt{3}}\left(
1-\frac{1}{216f_{0}}\left(  t-t_{0}\right)  ^{2}-\frac{1+9f_{0}\Lambda
}{19440f_{0}^{2}}\left(  t-t_{0}\right)  ^{4}+...\right)  .
\end{equation}
from which it is obvious the the contribution of the cosmological constant is
significant far from the movable singularity.

\section{Conclusions}

\label{sec5}

In this study we investigated the existence of exact and analytic solutions
for the modified teleparallel $f\left(  T\right)  $ gravitational theory in a
FLRW background space with nonzero spatial curvature. For the $f\left(
T\right)  $ function we assumed the proposed power-law models. We found that
there exists a movable singularity for the field equations which describes a
singular scaling solution for the field equations. The asymptotic behaviour
for the scale factor is that of a Milne-like scale factor, for negative and
positive spatial curvature for the background space. These generalized Milne
solutions are the analytic solutions for the power-law $f\left(  T\right)
=T^{n}$ theory.

However, for the $f\left(  T\right)  =T+f_{0}T^{n}-2\Lambda$ model, for the
derivation of the analytic solution, because of the existence of the movable
singularity, we applied the singularity analysis and specifically the ARS
algorithm. We proved that the modified Friedmann equations possess the
Painlev\'{e} property and the analytic solution for the scalar factor are
expressed in Right Puiseux Series, where the first term is the generalized
Milne behaviour. Because the Puiseux solutions are Right, that is, as far as
we move from the initial singularity the scale factor increases, consequently,
we can easily derive the attractor for the field equations and the solution
for large values of the independent parameter. In addition, the presence of
the cosmological constant term plays no role in the general evolution for the
dynamical system.

For large values of a nonconstant scale factor, from (\ref{fc.10}) it follows
$T\simeq-6H^{2}$ , that is, equation (\ref{fc.16}) becomes
\begin{equation}
0\simeq H^{2}\left(  -2\left(  1+nf_{0}\left(  -6H^{2}\right)  ^{n-1}\right)
-1+f_{0}\left(  -6H^{2}\right)  ^{n-1}\right)  .
\end{equation}
Hence, for it follows $H\left(  a\right)  =-6\left(  \frac{1}{f_{0}\left(
1-2n\right)  }\right)  ^{\frac{1}{n-1}}$, for $n\neq\frac{1}{2}$.\ Thus, this
model leads to an expanding universe, as we found from the singularity
analysis without necessary to consider a cosmological constant term.

It is important to mention here that the limit of General Relativity is not
recovered for a nonlinear function $f\left(  T\right)  $. Indeed, General
Relativity is recovered in $f\left(  T\right)  $ gravity, when it holds
$f\left(  T\right)  _{|T\rightarrow0}=0$, and $Tf^{\prime}\left(  T\right)
_{|T\rightarrow0}=0$. Thus, from (\ref{fc.10}) it is clear that the General
Relativity vacuum solution does not hold when $T=0.$

We found that in the presence of the spatial curvature in $f\left(  T\right)
$ gravity the universe becomes inflationary, while a generalized Milne-like
exact solution was found in the power-law model. In a future study we plan to
consider the asymptotic scale factor near to the movable singularity as a toy
model for the cosmological observations with a background space with spatial curvature.

\begin{acknowledgments}
This work is based on the research supported in part by the National Research
Foundation of South Africa (Grant Numbers 131604).
\end{acknowledgments}

\end{document}